\documentclass[10pt,english]{elsarticle}
\usepackage[T1]{fontenc}
\usepackage[latin9]{inputenc}
\usepackage{amssymb}

\makeatletter

\providecommand{\tabularnewline}{\\}

\usepackage{url}
\usepackage{tabularx}
\usepackage{listings}
\usepackage{xcolor}

\definecolor{mygreen}{rgb}{0,0.6,0}
\definecolor{mygray}{rgb}{0.5,0.5,0.5}
\definecolor{mymauve}{rgb}{0.58,0,0.82}

\makeatother

\usepackage{babel}
\usepackage{listings}

\begin{document}

\title{High performance computing aspects of a dimension independent semi-Lagrangian discontinuous Galerkin code\tnoteref{label1}} \tnotetext[label1]{This work is  supported by the Austrian Science Fund (FWF) -- project id: P25346.
The computational results presented have been achieved using the Vienna Scientific Cluster (VSC).}
\author[uibk]{Lukas Einkemmer\corref{cor1}} \ead{lukas.einkemmer@uibk.ac.at}
\address[uibk]{Department of Mathematics, University of Innsbruck, Austria}
\cortext[cor1]{Corresponding author}
\begin{abstract}
The recently developed semi-Lagrangian discontinuous Galerkin approach is used to discretize hyperbolic partial differential equations (usually first order equations). Since these methods are conservative, local in space, and able to limit numerical diffusion, they are considered a promising alternative to more traditional semi-Lagrangian schemes (which are usually based on polynomial or spline interpolation). 

In this paper, we consider a parallel implementation of a semi-Lagrangian discontinuous Galerkin method for distributed memory systems (so-called clusters). Both strong and weak scaling studies are performed on the Vienna Scientific Cluster 2 (VSC-2). In the case of weak scaling, up to 8192 cores, we observe a parallel efficiency above 0.89 for both two and four dimensional problems. Strong scaling results show good scalability to at least 1024 cores (we consider problems that can be run on a single processor in reasonable time). In addition, we study the scaling of a two dimensional Vlasov--Poisson solver that is implemented using the framework provided. All of the simulation are conducted in the context of worst case communication overhead; i.e., in a setting where the CFL number increases linearly with the problem size.

The framework introduced in this paper facilitates a dimension independent implementation (based on C++ templates) of scientific codes using both an MPI and a hybrid approach to parallelization. We describe the essential ingredients of our implementation.
\end{abstract} 
\maketitle

\section{Introduction}

The so-called semi-Lagrangian methods constitute a class of numerical
schemes used to discretize hyperbolic partial differential equations
(usually first order equations). The basic idea is to follow the characteristics
backward in time. Depending on the equation under consideration, this
can be accomplished either by using an analytic formula or an ODE
(ordinary differential equations) solver. Note, however, that since
the endpoint of a characteristic curve starting at a given grid point
usually does not coincide with the grid used, an interpolation procedure
has to be employed. An obvious choice is to reconstruct the desired
function by polynomial interpolation. However, piecewise polynomial
splines are usually preferred.

Note that, such interpolation procedures suffer from a number of shortcomings,
two of which we will now discuss in some detail. First, in the case
of polynomial interpolations not even conservation of mass is guaranteed
(although conservative semi-Lagrangian methods in the context of a
finite volume approximation have been considered; see \cite{crouseilles2010}).
The traditional approach of dealing with this shortcoming is to employ
a high order reconstruction. However, and this brings us to the second
shortcoming, such high order reconstructions have to employ large
stencils. While the computation of these stencils is usually cheap
on a single processor, they incur significant communication overhead
when implemented on distributed memory systems. In the case of spline
approximations, which according to Sonnendr\"ucker \cite{sonnendruecker2011}
are still considered the de facto standard in Vlasov simulations,
a sparse linear system of equations has to be solved (an algorithm
with low flop/byte ratio and significant communication overhead). 

On the other hand the semi-Lagrangian discontinuous Galerkin methods
employ a piecewise polynomial approximation in each cell of the computational
domain (see, for example, \cite{qiu2011}, \cite{crouseilles2011}
and \cite{einkemmer2014}). In case of a simple advection equation
the represented function is translated and then projected back to
the appropriate subspace of piecewise polynomial functions. These
methods, per construction, are mass conservative and only access two
adjacent cells in order to compute the necessary projection (this
is true independent of the order of the approximation). For low CFL
numbers (see section \ref{sec:method} for details) this results in
a comparable communication overhead as is needed for a low order polynomial
interpolation. However, linear and quadratic piecewise polynomials
(without a continuity constraint) show a comparable numerical diffusion
as a $9$th and $17$th order Lagrange interpolation, respectively.
If high frequencies are important, as is usually the case for Vlasov
simulations, the piecewise linear interpolation is even comparable
to the 17th order Lagrange interpolation (see \cite{crouseilles2011}).
This can result in a significant reduction in communication overhead
(especially for low CFL numbers). Furthermore, the semi-Lagrangian
discontinuous Galerkin methods compare very favorable with spline
interpolation. However, a direct comparison of the communication overhead
in this situation is more difficult as some variant of domain decomposition
can be employed in order to improve the scaling of the spline interpolation
(see, for example, \cite{crouseilles2009}). Note, however, that even
in such a domain decomposition approach a relatively large communication
overhead is incurred. This is due to the fact that the boundary condition
for the local spline reconstruction requires a large stencil if the
desirable properties of the global cubic spline interpolation are
to be preserved (the method derived in \cite{crouseilles2009} requires
a centered stencil of size $21$). In addition, the error propagation
of the semi-Lagrangian discontinuous Galerkin method, in case of the
advection equation, has been investigated and found to be superior
to both spline and Lagrange interpolation (see \cite{steiner2013}
and \cite{einkemmer2014b}).

A disadvantage of semi-Lagrangian discontinuous Galerkin methods is
that the memory scales linearly in the order of the method. However,
since in many applications a relatively low order (compared to other
semi-Lagrangian methods) is sufficient and no intermediate coefficients
have to be stored in memory (which is the case for the spline interpolation,
for example), they constitute a promising numerical method on present
and future high performance computing (HPC) systems. Let us note that
while well developed numerical software exists for semi-Lagrangian
methods employing polynomial and spline interpolation, e.g.~the GYSELA
code \cite{gysela} which scales to at least $65000$ cores on a Blue
Gene/Q system, this is not the case for the discontinuous Galerkin
approach considered here

It is therefore our intention to develop a software framework which
can be used to implement scientific computer codes based on a semi-Lagrangian
discontinuous Galerkin approach. In this paper, we present a dimension
independent parallelized implementation of a semi-Lagrangian discontinuous
Galerkin discretization (in C++ using MPI and OpenMP). In order to
manage code complexity, we will provide a number of dimension independent
functions (based on C++ templates) that facilitate the construction
of numerical solvers of arbitrary dimension (see section \ref{sec:computer-code}).
In addition, we analyze weak (problem size is inreased linearly with
the number of cores) and strong scaling (problem size is constant
as the number of cores increases) behavior for two and four dimensional
problems on the VSC-2 HPC system%
\footnote{The VSC-2 (Vienna Scientific Cluster 2) consists of 21024 Opteron
Magny Cours 6132HE cores with 2 GB DDR3 memory per core (32 GB per
node) and an Infiniband QDR interconnect. See \url{vsc.ac.at} for
more information. Due to scheduling limitations we have limited our
simulations to a maximum of $8192$ cores.%
} (see section \ref{sec:Numerical-results}). Furthermore, we briefly
discuss a hybrid parallelization strategy (using OpenMP on each socket
and MPI between sockets). As an application, we present weak scaling
results for a Vlasov--Poisson solver. The Vlasov--Poisson equations
are widely used in plasma physics.

\section{The semi-Lagrangian discontinuous Galerkin method\label{sec:method}}

In this section we describe the semi-Lagrangian discontinuous Galerkin
method and its performance characteristics. For simplicity we will
limit ourselves to the initial value problem for the advection equation
given by 
\begin{equation}
\partial_{t}u(t,x)+v\partial_{x}u(t,x)=0,\qquad u(0,x)=u_{0}(x),\label{eq:advection}
\end{equation}
where $u$ is the sought after function whereas $v\in\mathbb{R}$
as well as $u_{0}$ are given. For a more detailed treatment we refer
the reader to \cite{crouseilles2011} and \cite{crouseilles2012}.
The characteristics of equation (\ref{eq:advection}) can be derived
analytically. Therefore, we can write down the exact solution as follows
\[
u(t,x)=u_{0}(x-vt).
\]
Note that the solution of this problem is the fundamental building
block for a splitting approach to the Vlasov--Poisson and Vlasov--Maxwell
equations. It can also be used to handle a Burgers' type nonlinearity
(see \cite{einkemmer2014c}).

Now, let us consider the discretization of this equation within the
semi-Lagrangian discontinuous Galerkin approach. We divide our domain
in cells $C_{i}$ and assume that a function $\tilde{u}$ is given
such that $\tilde{u}\vert_{C_{i}}$ , i.e. the restriction of $\tilde{u}$
to the $i$th cell, is a polynomial of degree $k$. Then the function
$\tilde{u}$ lies in the approximation space. However, the translated
function $x\mapsto\tilde{u}(x-v\tau)$, where $\tau$ is the time
step size, can, in general, not be represented in this form. Therefore,
we use a projection operator $P$ and compute (numerically)
\[
\tilde{u}^{n+1}=P\tilde{u}^{n}(\cdot-vt).
\]
The function $\tilde{u}^{n+1}$ constitutes the sought after approximation
of $\tilde{u}(x-v\tau)$. The operator $P$ is the $L^{2}$ projection
on the (finite dimensional) subspace of cell-wise polynomials of degree
$k$.

To determine the numerical method we have to chose a basis of the
finite dimensional subspace. This then also determines the degrees
of freedom stored in computer memory. In \cite{einkemmer2014} the
basis of Legendre polynomials up to degree $k$ has been employed.
The projection can then be computed either analytically or by Gaussian
quadrature. On the other hand in \cite{crouseilles2011} the degrees
of freedom are chosen as the Gauss--Legendre points in each cell (this
corresponds to a basis of certain Lagrange basis polynomials). The
projection is then computed by Gaussian quadrature. The numerical
effort of both implementations is comparable. In the implementation
considered here we employ the latter approach. This is due to the
fact that the degrees of freedom correspond to function evaluations
at given points. This makes the extension to arbitrary dimensions
easier and facilitates the interoperability with third party libraries.
The explicit formulas necessary for the implementation are derived
and stated in \cite{crouseilles2011}. 

To conclude this section, let us discuss the computational effort
and communication overhead of the semi-Lagrangian discontinuous Galerkin
method considered here. Counting the number of arithmetic operations
for every of the $n^{d}o^{d}$ degrees of freedom, where $n$ is the
number of cells, $o=k+1$ is the order of the polynomial approximation,
and $d$ is the space dimension, we obtain a total of  $16o^{3}+15o^{2}+7o+3$
floating point operations per time step. For piecewise constant, linear,
and quadratic polynomials the flops per step and the flops per byte
(read from main memory) are shown in Table \ref{tab:flops-per-step}.

\begin{table}
\begin{centering}
\begin{tabular}{ccc}
\hline 
$o$ & flops/step & flops/byte\tabularnewline
\hline 
$1$ & $41n^{d}o{}^{d}$ & $2.6$\tabularnewline
$2$ & $205n^{d}o{}^{d}$ & $12.8$\tabularnewline
$3$ & $591n^{d}o{}^{d}$ & $36.9$\tabularnewline
\end{tabular}
\par\end{centering}

\protect\caption{Number of floating point operation per time step are shown for the
semi-Lagrangian discontinuous Galerkin scheme considered in this paper.
In addition, the flop to byte ratio for double precision floating
point numbers is displayed in terms of the number of cells $n$ and
the order of the approximation in each cell $o$. If single precision
arithmetics is used, the flop/byte ratio is halved. \label{tab:flops-per-step}}
\end{table}
Let us note that the problem under consideration is compute bound
on most modern architectures for polynomials of degree one or above.
The implementation is currently not aggressively optimized, in order
to maintain code readability, and no effort has been made to vectorize
the code (besides the vectorization that is performed by the compiler).

Let us now discuss the communication overhead necessary on a distributed
memory model. The scheme under consideration, as is true for semi-Lagrangian
methods in general, is not restricted by a CFL (Courant\textendash Friedrichs\textendash Lewy)
condition. In what follows we denote the CFL number%
\footnote{In case of the advection equation considered in this section, the
CFL number is defined as $\tau v$ divided by the cell size, where
$\tau$ is the time step size and $v$ the advection speed.%
} by $C$. Each MPI process communicates $n_{0}^{d-1}\lceil C\rceil o$
double precision floating points numbers to exactly one adjacent MPI
process, where $n_{0}^{d}$ is the number of floating point numbers
stored per process.

\section{Numerical results\label{sec:Numerical-results}}

In this section we provide weak and strong scaling results for the
advection problem introduced in section \ref{sec:method}. As the
metric of merit we use parallel efficiency, which is defined as the
(theoretical) run time of a program that scales ideally divided by
the measured run time. In addition, we discuss an implementation of
a Vlasov--Poisson solver (based on the discontinuous Galerkin semi-Lagrangian
framework discussed here). Before proceeding, let us note that the
communication overhead of our computer program is due to MPI communication,
preparation of the boundary data necessary for this communication,
and synchronization overhead. It is essential for an efficient implementation
that the MPI communication is interleaved with computation (to hide
the associated communication overhead). However, even if this is accomplished,
as is the case for most weak scaling studies conducted in this paper,
the parallel efficiency is influenced by the two other factors as
well.

First, let us conduct a weak scaling study of the two dimensional
advection equation. Since for a two dimensional problem we can run
numerical simulations of appreciable size (at least up to $1024$
cells in each direction) on a single node, we will consider both weak
and strong scaling here. In all the simulations conducted we fix the
time step size. Therefore, the CFL number, and thus the amount of
data communicated by any two processors increases linearly with the
problem size. Let us note that this configuration constitutes the
worst case as far as communication overhead is concerned. In a realistic
simulation, however, we would most likely have to decrease the time
step size simultaneously with increasing the problem size in order
to keep the time integration error sufficiently small. 

The weak scaling results are given in Table \ref{tab:weakscaling-advection}.
We observe excellent scaling for up to $8192$ cores. Furthermore,
the timing results provided by the computer program do suggest that
we are able to hide the MPI communication overhead almost perfectly
up to the maximum number of cores considered here. Thus, we would
expect that this implementation scales to larger HPC systems as well.

\begin{table}
\begin{minipage}[t]{0.49\linewidth} \centering
\begin{tabular}[t]{ l r r } 
\hline
\# cores & time & efficiency \\
\hline
16 	 & 14.36 	 & 1.00 	 \\
64 	 & 14.80 	 & 0.97 	 \\
256 	 & 14.88 	 & 0.96 	 \\
1024 	 & 15.28 	 & 0.94 	 \\
2048 	 & 15.23 	 & 0.94 	 \\
4096 	 & 15.26 	 & 0.94 	 \\
\end{tabular}
\end{minipage} \begin{minipage}[t]{0.49\linewidth} \centering
\begin{tabular}[t]{ l r r } 
\hline
\# cores & time & efficiency \\
\hline
16 	 & 14.90 	 & 1.00 	 \\
64 	 & 15.34 	 & 0.97 	 \\
256 	 & 15.03 	 & 0.99 	 \\
1024 	 & 15.37 	 & 0.97 	 \\
2048 	 & 15.71 	 & 0.95 	 \\
4096 	 & 15.58 	 & 0.96 	 \\
8192 	 & 16.09 	 & 0.93 	 \\ 
\end{tabular}
\end{minipage}

\protect\caption{Weak scaling for the two dimensional advection equation using $512\times512$
cells (left) and $1024\times1024$ cells (right) with piecewise polynomials
of degree $1$ ($o=2$) per MPI process. The run time as well as the
parallel efficiency, compared to a single node, is shown (for ideal
scaling the measured run time would remain constant). The CFL number
for $512\times512$/$1024\times1024$ cells per MPI process ranges
from approximately $0.5$/$1$ for a single core to approximately
$32$/$91$ for $4096$/$8192$ cores.\label{tab:weakscaling-advection}}
\end{table}
Now, let us consider the strong scaling results given in Table \ref{tab:strongscaling-advection}.
In this case the communication overhead increases with the number
of cores while the computation time decreases. Therefore, starting
at a certain number of cores (depending on the problem size) we are
not able to hide the communication overhead anymore and the parallel
efficiency degrades significantly. Let us further note that the choice
of the CFL number has a significant influence on the strong scaling
that is achieved in such simulations. This is due to the fact that
the communication overhead increases linearly with the CFL number.

\begin{table}
\begin{minipage}[t]{0.49\linewidth} \centering
\begin{tabular}[t]{ l r r } 
\hline
\# cores & time & efficiency \\
\hline
16 	 & 59.49 	 & 1.00 	 \\
64 	 & 16.47 	 & 0.90 	 \\
256 	& 4.71 	  & 0.79 	 \\
1024    & 1.80 	  & 0.52 	 \\ 
\end{tabular}
\end{minipage} \begin{minipage}[t]{0.49\linewidth} \centering
\begin{tabular}[t]{ l r r } 
\hline
\# cores & time & efficiency \\
\hline
16 	 & 51.47 	& 1.00    \\
64 	 & 12.93 	& 1.00    \\
256 	& 3.40 	 & 0.95    \\
1024    & 0.98 	 & 0.82    \\
2048    & 0.55    & 0.73 	 \\
\end{tabular}
\end{minipage}

\protect\caption{Strong scaling for the two dimensional advection equation using a
total of $256\times256$ cells with a CFL number of approximately
$1$ (left) and $1024\times1024$ cells with a CLF number of approximately
$5$ (right). In both problems piecewise polynomials of degree $1$
($o=2$) are employed. The run time as well as the parallel efficiency
is shown (for ideal scaling the measured run time would decrease linearly
with the number of cores used).\textbf{ \label{tab:strongscaling-advection}}}
\end{table}

Second, let us consider the four dimensional advection equation. In
this case we can not run a problem of appreciable size on a single
core. Thus, we will only consider weak scaling here. The corresponding
results are given in Table \ref{tab:4d-weakscaling-advection}. Let
us make two observations. First, while we observe good scaling up
to $8192$ cores, the parallel efficiency is lower than in the case
of the two dimensional advection. This, however, is expected as the
ratio of computation to communication overhead is significantly lower
in the four dimensional problem (as compared to the two dimensional
problem considered above). 

\begin{table}
\begin{minipage}[t]{0.49\linewidth} \centering
\begin{tabular}[t]{ l r r } 
\hline
\# cores & time & efficiency \\
\hline
16 	 & 6.00 	 & 1.00 	 \\
256 	 & 6.23 	 & 0.96 	 \\
1024 	 & 6.32 	 & 0.95 	 \\
2048 	 & 6.37 	 & 0.94 	 \\
4096 	 & 6.71 	 & 0.89 	 \\
8192 	 & 6.89 	 & 0.87 \\
\end{tabular}
\end{minipage} \begin{minipage}[t]{0.49\linewidth} \centering
\begin{tabular}[t]{ l r r } 
\hline
\# cores & time & efficiency \\
\hline
16 	 & 12.50 	 & 1.00 	 \\
256 	 & 13.78 	 & 0.91 	 \\
1024 	 & 13.89 	 & 0.90 	 \\
2048 	 & 15.12 	 & 0.83 	 \\
4096 	 & 15.65 	 & 0.80 	 \\
\end{tabular}
\end{minipage}

\protect\caption{Weak scaling for the four dimensional advection equation using $16^{4}$
cells (left) and $32^{4}$ cells (right) with piecewise polynomials
of degree $1$ ($o=2$) per MPI process. The run time as well as the
parallel efficiency, compared to a single node, is shown (for ideal
scaling the measured run time would remain constant). For $16^{4}$/$32^{4}$
cells per MPI process the CFL number ranges from approximately $0.6$/$1.3$
for a single core to approximately $6$/$10$ for $8192$/$4096$
cores.\textbf{\label{tab:4d-weakscaling-advection}}}
\end{table}

\begin{figure}
\begin{minipage}[t]{0.44\linewidth} \centering
\centering \vspace{0pt}
\begin{tabular}{ l r r } 
\hline
\# cores & time & scaling \\
\hline
16 	 & 21.18 	 & 1.00 	 \\
64 	 & 22.43 	 & 0.94 	 \\
256 	 & 23.70 	 & 0.89 	 \\
1024 	 & 24.03 	 & 0.88 	 \\
2048 	 & 24.31 	 & 0.87 	 \\
4096 	 & 24.76 	 & 0.86 	 \\
8192 	 & 24.30 	 & 0.87
\end{tabular}
\end{minipage} \begin{minipage}[t]{0.55\linewidth} \centering
\vspace{0pt}
\includegraphics[width=0.99\textwidth]{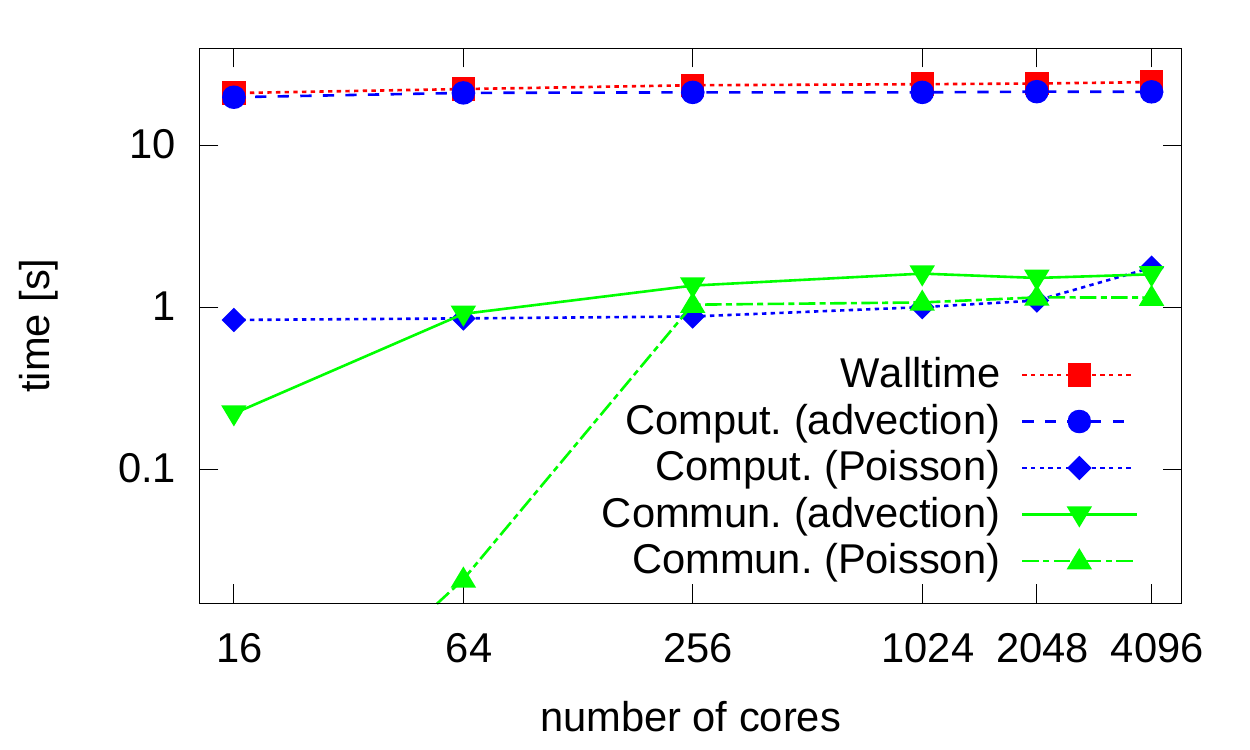}
\end{minipage}

\protect\caption{Weak scaling for the two dimensional Vlasov--Poisson equation using
$512\times512$ cells ($n=512$) and piecewise polynomials of degree
$1$ ($o=2$) per MPI process. The run time as well as the scaling
behavior is shown (for ideal scaling the measured run time would remain
constant). The time step size is chosen as $\tau=0.01$. This results
in a worst case CFL number of approximately $2.5$ for a single core
and approximately $221$ for \textbf{$8192$} cores. The plot on the
right shows the run time (of the different parts of the program) as
a function of the number of cores.\textbf{ \label{fig:weakscaling-vlasov}}}
\end{figure}

Third, let us consider the Vlasov--Poisson equations as an example
of an application of the semi-Lagrangian discontinuous Galerkin framework
discussed in this paper. We employ the numerical approach outlined
in \cite{einkemmer2014} and use the FFTW \cite{FFTW05} library to
compute a solution to Poisson's equation. Once again, we employ a
fixed time step size which results in a CFL number that scales linearly
with the problem size. The weak scaling results for this configuration
are shown in Figure \ref{fig:weakscaling-vlasov}. In the plot on
the right we show the run time of the different parts of the program
as a function of the number of cores used. Note that while the computational
effort of Poisson's equation is negligible (as is widely recognized
in the literature as Poisson's equation is posed in a single space
dimension only), the communication time necessary for the three advections
is similar to that of the fast Fourier transform. This additional
communication overhead, due to the collective communication necessary
to solve Poisson's equations, accounts for the slight additional decrease
in parallel efficiency as compared to the advection problem considered
in Table \ref{tab:weakscaling-advection}. Furthermore, let us note
that a more detailed analysis reveals that the communication for the
three advections is almost completely hidden by the corresponding
computation. The measured time (as shown in the plot) is mainly due
to the preparation of the boundary data (i.e., populating the proper
data structure) and synchronization overhead.

To complete this section, let us briefly discuss hybrid programming.
All the simulations above are conducted using MPI for parallelization.
A hybrid approach complements this by using OpenMP on each node and
MPI to communicate between different nodes. Such a hybrid implementation
has the (potential) advantage that less communication overhead is
incurred (each node represents a shared memory architecture). We have
implemented this hybrid programming model for the simulations conducted
above. However, up to the number of cores considered here, we have
not observed a significant increase in performance compared to the
MPI only implementation. Let us note, however, that in case of simulations
which have a very large worst case CFL number (such as for initial
values of the Vlasov--Poisson equations that decay only algebraically)
it enables us to consider larger problem sizes while still maintaining
a communication scheme where each MPI process exclusively talks with
its neighbors. In addition, a slight decrease in the total amount
of memory consumed can be observed for the hybrid implementation (up
to approximately 15\%).

\section{Description of the computer code\label{sec:computer-code}}

In this section we will describe the three main aspects of the implementation
necessary to extend existing codes (such as the advection and Vlasov--Poisson
implementation used in the previous section) and to implement new
schemes based on the semi-Lagrangian discontinuous Galerkin framework
described here.

The dimension of the problem is a template parameter denoted by \texttt{d}.
Thus, in order to change the dimension of a given implementation we
have to recompile the application. This, in our view, is not a serious
limitation; furthermore, it enables the compiler to apply additional
optimizations for each use case. Also, in this framework it is very
simple to supply specific optimizations which can only be applied
to a two dimensional problem, for example. As the index type we use
the typesafe \texttt{boost::array} class from the Boost library.

In developing dimension independent scientific codes it is vital to
have a construct that enables us to loop over all (or a subset) of
the indices in a multi-dimensional array. Our framework provides the
function \texttt{iter\_next} to facilitate this behavior. Its prototype
and a usage example are given in the following listing.

\clearpage
\begin{lstlisting}[language={C++}]
template<size_t d> 
bool iter_next(array<int,d> &idx, array<int,d> max,
	array<int,d>* start = NULL);

array<int, d> idx; array<int,d> max; 
// initialization of idx and max
do { 	
	// computation
} while(iter_next(idx,max)); 
\end{lstlisting}
The argument \texttt{idx} specifies the current (multi-)index and
\texttt{max} denotes the upper bound of the iteration in each direction.
In addition, the optional third parameter can be used to specify a
starting value that is different from zero in one or more directions.

In order to perform communicate over the MPI interface the appropriate
slice of the multi-dimensional data has to be extracted. To facilitate
this, in a dimension independent way, the following two functions
are provided.

\begin{lstlisting}[language={C++}]
void pack_boundary(int dim, leftright lr, int bdr_len,
	vector<double>& bdr_data, 
	vector<double>* lalpha=NULL, int dim2=0);
double* unpack_boundary(array<int,2*d-2> idx, int dim, 
	int bdr_len, vector<double>& bdr_data)
\end{lstlisting}
The user has to specify the dimension in which the translation is
to be conducted (\texttt{dim}), the number of cells that need to be
communicated (in the \texttt{dim}-direction), and a vector to hold
the boundary data (\texttt{bdr\_data}). In addition, the direction
of the translation is given (\texttt{lr}); alternatively, also a vector
of advection speeds can be specified (\texttt{lalpha}). In the latter
case the direction of the translation can be different at each degree
of freedom (in the \texttt{dim2}-direction). The vector \texttt{bdr\_data}
is then send via an appropriate MPI call. The receiving process uses
the \texttt{unpack\_boundary} function in order to determine the appropriate
pointer that is passed to the \texttt{translate1d} function. 

The translate1d function performs the actual computation of the advection
and its prototype is given in the following listing.

\begin{lstlisting}[language={C++}]
void translate1d(typename node::view& u, 
	typename node::view& out, double* boundary,
	int boundary_size, double dist,
	translate1d_mode tm=TM_ALL);
\end{lstlisting}
The first two arguments specify the input and output iterator, respectively.
The pointer to the appropriate boundary data is determined by the
\texttt{unpack\_boundary} function. In addition, we have to specify
the number of cells in the translation direction (\texttt{boundary\_size}),
and the distance (\texttt{dist}) of the translation. In addition,
for the last argument we can specify \texttt{TM\_INTERIOR}. In this
case only the part of the computation that can be done without knowing
\texttt{boundary} is performed (in this situation it is permissible
to pass a \texttt{NULL} pointer as the third argument). This then
has to be followed up, in order to complete the computation, by an
additional call which specifies \texttt{TM\_BOUNDARY} as the last
argument and a proper pointer to boundary data. This is useful for
interleaving communication with computation and thus has been applied
in all the numerical simulations conducted in section \ref{sec:Numerical-results}.

The source code can be build using the included Makefile; it generates
an executable named \texttt{sldg} that can be executed as follows
\begin{lstlisting}[language=bash]
./sldg identifier
\end{lstlisting}
where identifier is one of \texttt{test\_1d\_order}, \texttt{test\_2d\_order}
(checks the order of the approximation for a simple advection), \texttt{test\_1d\_mpi},
\texttt{test\_2d\_mpi}, \texttt{test\_2d\_hybrid}, \texttt{test\_4d\_mpi},
\texttt{test\_4d\_hybrid} (solves the advection equation in parallel),
and \texttt{vlasovpoisson} (solves the Vlasov--Poisson equations in
parallel). Further options can be specified. A overview of available
options is displayed by running

\begin{lstlisting}[language=bash]
./sldg --help
\end{lstlisting}
The current version of the software can be obtained from \url{https://bitbucket.org/leinkemmer/sldg}.

\section{Conclusion \& Outlook}

The results presented in this paper show that the semi-Lagrangian
discontinuous Galerkin approach can be parallelized efficiently to
at least $8192$ cores on a modern HPC system. It is our believe that
the framework introduced here greatly facilitates the implementation
of scientific computer codes based on this discretization. As future
work we will consider two extensions of the implementation presented
here.

First, many modern HPC systems include up to a few hundred thousand
cores. Therefore, to investigate and optimize scaling to larger HPC
systems, compared to what we have considered in this paper, is certainly
of interest. This would also allow us to treat five or even six dimensional
problems (which is of interest in the study of kinetic plasma models,
for example). In addition, improving the single threaded performance
by vectorization would conceivably lead to significant gains in performance.

Second, in this paper we have only considered briefly the application
to problems of practical interest. In this context, we are mostly
interested in kinetic models of plasma physics (such as the Vlasov--Poisson
equations, Vlasov--Maxwell equations, and gyrokinetic models) and
in advection dominated second order partial differential equations.
In the latter case a splitting approach has to be employed, which
allows us to treat the diffusion term separately.

\bibliographystyle{plain}
\bibliography{sldg}

\end{document}